\documentclass[smallextended]{svjour3}

\usepackage{graphics,subfigure,amssymb,mhequ}

\journalname{Journal of Statistical Physics}

\begin{document}
\title{Sub-exponential mixing of open systems with particle-disk interactions}
\author{Tatiana Yarmola
\thanks{supported by the ERC Advanced Grant "Bridges"} \and \thanks{Department of Theoretical Physics, University of Geneva, Ecole de Physique, 24 quai Ernest-Ansermet, 1211 Genève 4. Tel: +41(22)379-6393.}}                   
\institute{University of Geneva \email{tatiana.yarmola@unige.ch}}
\date{Received: date / Accepted: date}
%
%
\maketitle
\begin{abstract}
  We consider a class of mechanical particle systems with deterministic particle-disk interactions coupled to Gibbs heat reservoirs at possibly different temperatures. We show that there exists a unique (non-equilibrium) steady state. This steady state is mixing, but not exponentially mixing, and all initial distributions converge to it. In addition, for a class of initial distributions, the rates of converge to the steady state are sub-exponential.
\end{abstract}

\section{Introduction}
Rigorous derivations of heat conduction laws for mechanical particle models coupled to heat reservoirs remain a mathematical challenge. A variety of models have been introduced in the past \cite{Balint,Collet,Anharmonic_EPR,Nonequilibrium,Klages_Nicolis_Rateitschak,Larralde_Leyvraz_MejiaMonasterio,Lin}; nearly all of the proposed derivations of the Fourier Law are partial solutions based on unproven assumptions \cite{HarmonicFourier,HarmomicCrystal,Challenge,AnharmonicFourier,Nonequilibrium,WeakAnharmonicFourier}. Developing proofs of these assumptions would require deep understanding of the properties of systems in non-equilibrium, i.e., coupled to several unequal heat reservoirs. The standard assumptions include the existence of the unique invariant measure (steady state) as well as certain bounds on the rates of convergence of initial distributions to the invariant measure.

For systems in equilibrium, i.e., when the temperatures of all the reservoirs are the same, the steady states can often be written down explicitly. The question of existence of non-equilibrium steady states has been open for practically all mechanical particle systems, by which we mean Hamiltonian-like systems, driven by stochastic heat reservoirs. The main difficulty lies in dealing with the non-compactness of the phase space. For the systems under consideration, however, it is relatively easy to envision scenarios under which particles slow down (freezing) or speed up (heating) due to stochasticity of the heat reservoirs. This  may push initial distributions towards zero or infinite energy levels and ultimately violate existence of physically relevant steady states.

An example of freezing has been observed (numerically) in one of the proposed models \cite{Freezing}. In that model, a particle acquires very low values of the speed once in a while under the evolution of the dynamics (due to stochasticity). Low values of the speed lead to long traveling times between collisions during which the particle has no influence on the evolution of the system. It is observed for the system in \cite{Freezing} that more and more particles get stuck on the low energy states resulting in fewer and fewer collisions per unit time. To rule out such unfortunate scenarios one must be able to control the probabilities of particles acquiring low speeds and the rates at which the speeds recover to normal ranges.

The dynamics of the mechanical particle systems driven by heat reservoirs may be viewed as a continuous-time Markov process. Harris' ergodic theorem and its generalizations are common tools for obtaining existence and uniqueness of the steady states as well as (exponential) convergence of initial distributions to the steady state. In discrete time, the theorem requires two things: to produce a non-negative potential $V$ (Lyapunov function) on the phase space which, on average, decreases as a power law under the push forwards of the dynamics, and, given such a $V$, to show minorization or Doeblin's condition on certain level set of $V$. The first condition guarantees that the dynamics enters the 'center' of the phase space, a certain level sets of $V$, with good control on the rates; and, once at the 'center', coupling is guaranteed by the minorization condition. Harris' ergodic theorem was applied in \cite{Khanin} for a discretization of the original continuous-time Markov process. In \cite{Y2} the results were extended to continuous time. The existence of a unique steady state was obtained by constructing a suspension flow over the discrete dynamics; convergence of initial distributions to the steady state followed from a result in general state space Markov process theory \cite{MeynII} once irreducibility of time-$1$ discrete process was shown. In addition, \cite{Y2} demonstrated that this convergence is sub-exponential for a class of initial distributions. The slow rates of convergence are due to the abundance of slow particles in the system which, in turn, do not influence the system for extended periods of time. This slows the rates of mixing.

The analysis in \cite{Khanin,Y2} relies heavily on the fact that there exists a meaningful discretization of the system that mixes exponentially fast. Because particles do not interact, the study of the dynamics of one particle on the collision manifold reveals important dynamical properties that yield implications for the continuous-time system. For an interacting particle system, the rates of mixing for the continuous-time system and its relevant discretizations happen to be comparable due to the slow particle effect. In a collision map, for example, slow particles experience rare collisions, which slow mixing down.

Sub-exponential mixing seems to be prevalent for canonical interacting particle systems driven by Gibbs heat reservoirs. If a discrete system does not mix exponentially, finding a potential $V$ that would still guarantee existence of invariant measures is a very diligent task requiring extremely good understanding of the dynamics of the system. Thus, one needs different methods to tackle the question of existence of non-equilibrium steady states. A very clear, and useful exposition of ideas and difficulties associated with this task is presented in \cite{GLP}. This paper relies heavily on the paper by Meyn and Tweedie \cite{Meyn_existence}, which provides a general framework of showing existence of invariant probability measures for general state space Markov processes.

In this paper we consider a class of mechanical systems in which particles interact with an 'energy tank' represented by a rotating disk anchored at the center. Particles move freely between collisions with the tank or the boundary. When a particle collides with the disk, an energy exchange occurs, in which the particle exchanges the tangential component of its velocity with the angular velocity of the disk and the normal component of the particle's velocity changes sign. A system in this class is coupled to heat reservoirs set at possibly different temperatures that absorb particles when they collide with the boundaries of the reservoirs and emit new particles according to the Gibbs distribution corresponding to the temperatures of the reservoirs. Such systems were introduced in \cite{Klages_Nicolis_Rateitschak,Larralde_Leyvraz_MejiaMonasterio} and further studied in \cite{Nonequilibrium,Lin}. Even though introducing turning disks is a rather natural generalization of billiards, controlling the deterministic dynamics of such systems (event in the absence of heat reservoirs) is a delicate task and few results exist \cite{BalintRotor,Bunimovich}. Consequently, the result on our paper must rely on an additional simplification: it is provided by the frequent refreshing of particles at the heat reservoirs.

The main geometric assumption that makes our the analysis feasible is that a particle can hit the disk at most once before returning to the heat reservoir. The domain is assumed to be circular (see Fig. \ref{fig: the system}) as in \cite{BalintRotor}, which simplifies the analysis, but is not essential and can be generalized. In addition, we introduce two vertical walls to split the domain in two. The sole purpose of the walls is to create a visual image of separation between the two reservoirs: our results apply directly if the walls are removed. Our method only shows existence and mixing properties of the non-equilibrium steady state of the system and does not provide any description of the steady state itself. We leave the description of the steady state for the future work.

The existence of a non-equilibrium steady state is shown in section \ref{sect: existence} through estimating hitting times of a carefully chosen compact set $C$ in continuous time without an aid of a discretization or a potential. A regeneration times idea is employed in the argument. In order to apply a general state space Markov process theory developed in \cite{Meyn_existence}, one also needs to show that the minorization condition holds on $C$, which we do in \ref{subsect: C petite}. Convergence of initial distributions to the steady state follows by application of a theorem in \cite{MeynII} after a small modification of the minorization condition argument. In section \ref{sect: non exp mixing} we show that mixing is not exponential and for a large class of initial distribution convergence of initial distributions to the steady state occurs at sub-exponential rates. The argument is similar to \cite{Y2}. However, we cannot use the potential $V$ for certain upper bound estimates and different methods are required. The key property that leads to sub-exponential mixing is that the (invariant) measure of the states for which at least one particle will not collide with a heat reservoir or a disk in time $\tau$ is of the order of $\tau^{-2}$. The dynamics thus resembles one of an expanding map with a neutral fixed point and sub-exponential convergence rates for a class of initial distributions can be obtained using arguments similar to \cite{LSY}. The bounds on the measure of the particles that will not collide with a heat reservoir or a disk in time $\tau$ are obtained using the minorization condition and other properties of the dynamics.

\section{Settings} \label{sect: settings}

\begin{figure}
  \centering
  \subfigure[The configuration]{\resizebox{0.4\textwidth}{!}{
  \includegraphics{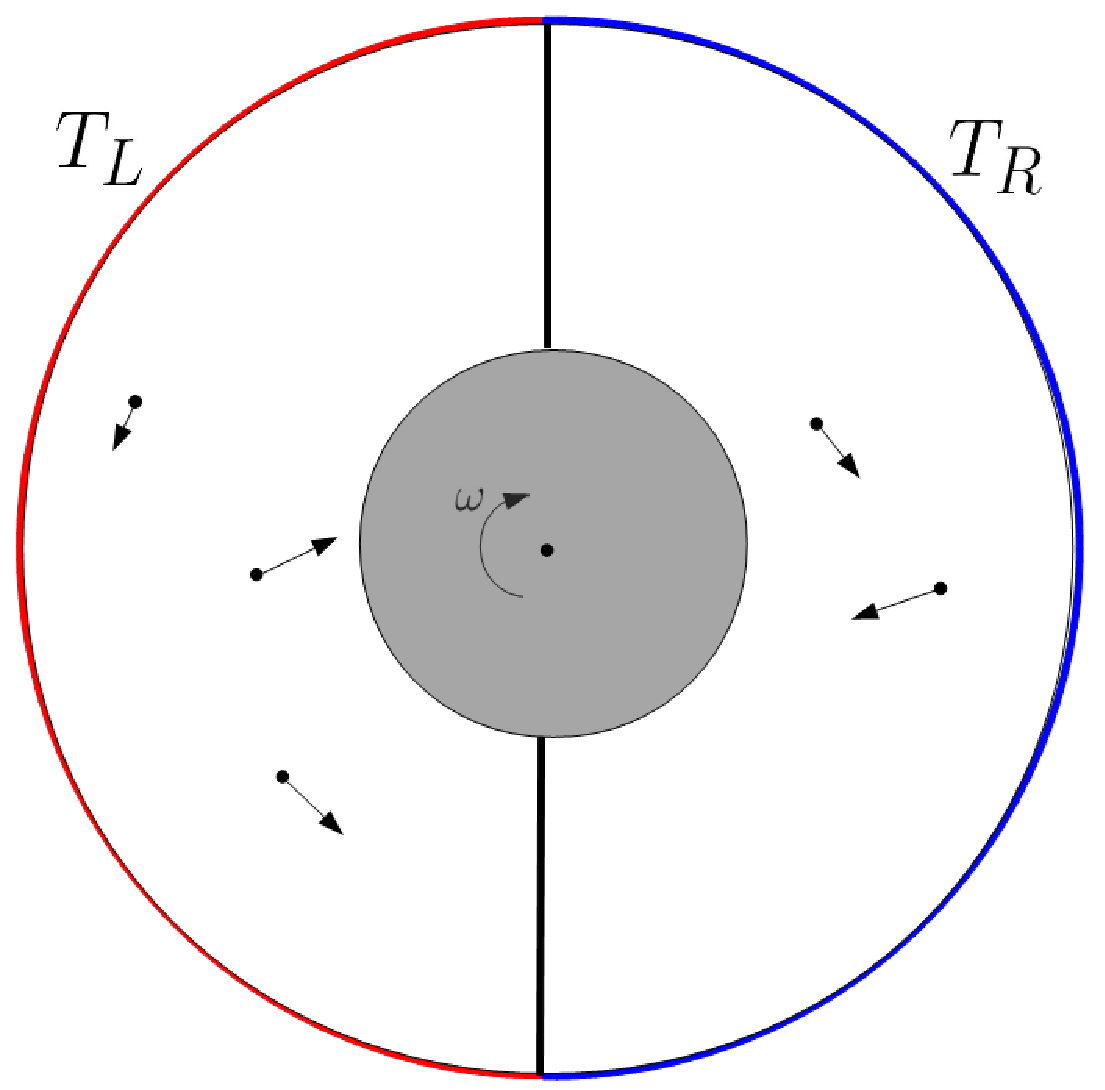}}  \label{fig: manyparticles} }
  \subfigure[Suspension flow coordinates]{\resizebox{0.44\textwidth}{!}{
  \includegraphics{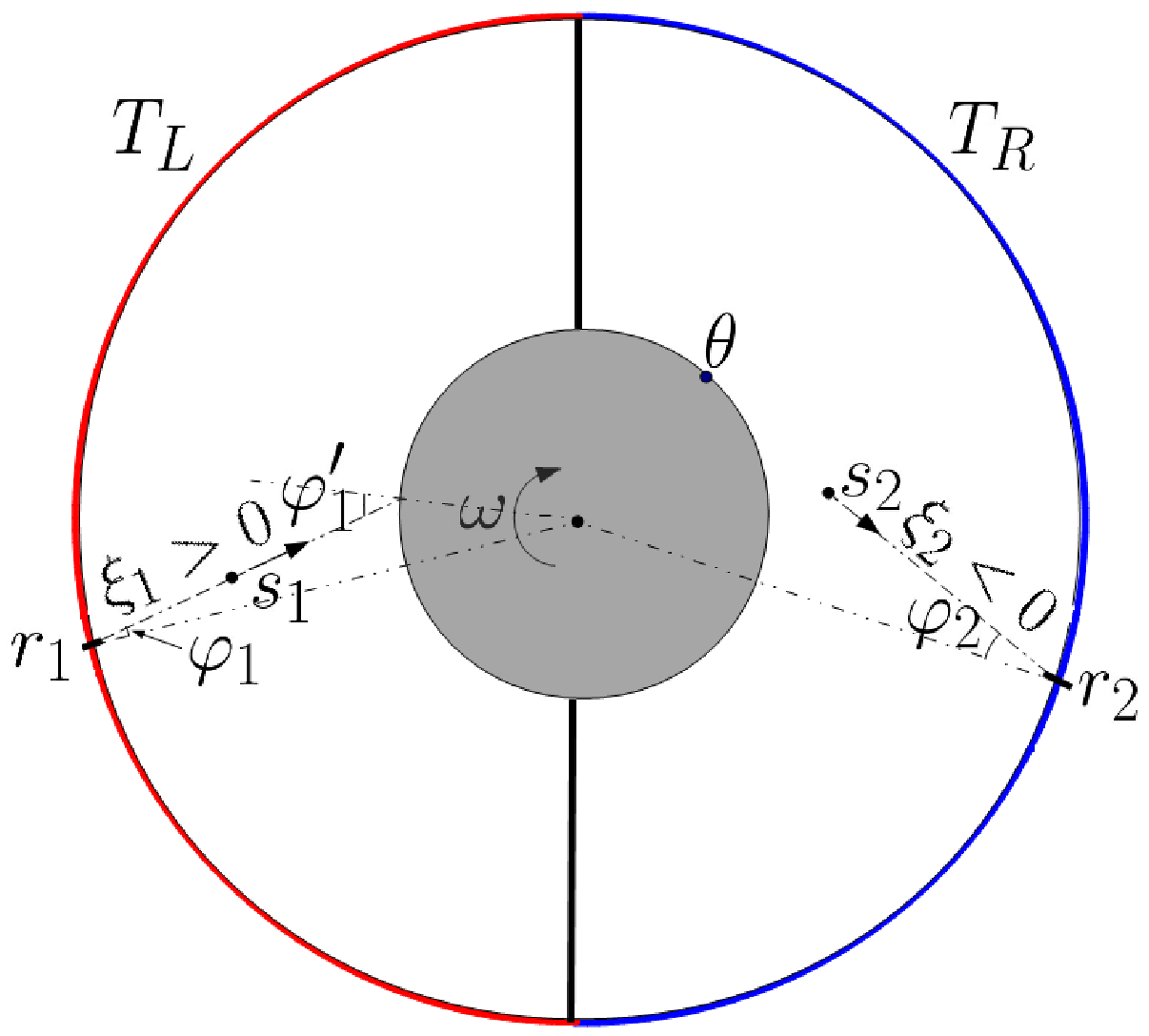}} \label{fig: coordinates} }
  \caption{The system}\label{fig: the system}
\end{figure}

Let $\Gamma$ be a circular domain of radius $R+d$. A disk $D$ is anchored at the center of $\partial \Gamma$. The disk $D$ is allowed to rotate freely with angular velocity $\omega$; denote by $\theta \in S^1$ a position of a marked point on $\partial D$. Let $L_u$ and $L_d$ be two vertical walls on the top and on the bottom of $D$, splitting $\Gamma \setminus D$ into two halves, left and right. See Fig.~\ref{fig: manyparticles}. Our system consists of $k$ particles in $\Gamma \setminus D$ with positions $(x_1, x_2, \cdots x_k)=:\mathbf{x}$ and velocities $(v_1,v_2, \cdots, v_k)=:\mathbf{v}$. The particles are confined to either of the two halves and move freely between collisions with $\partial \Gamma \cup \partial D \cup L_u \cup L_d$. The collisions with $L_u$ and $L_d$ are specular, i.e., the angles of incidence are equal to the angles of reflection. When a particle collides with the boundary of the disk $\partial D$, an energy exchange occurs, in which the particle exchanges the tangential component of its velocity with the angular velocity of the disk and the normal component of the particle's velocity changes sign. More precisely, if $v=(v_t,v_\perp)$ is the particle's velocity decomposition upon collision with $\partial D$ and the disk rotates with angular velocity $\omega$, then the post-collision velocities are:
$$v_\perp' = - v_\perp, \;\; v_t'=\omega, \;\; \mathrm{and} \;\; \omega'=v_t,$$
where $v'=(v_\perp',v_t')$ is the particles velocity decomposition and $\omega'$ is the disk's angular velocity immediately after the collision. This interaction was introduced in \cite{Klages_Nicolis_Rateitschak,Larralde_Leyvraz_MejiaMonasterio}.

The left and right parts of $\partial \Gamma$, $\partial \Gamma_1$ and $\partial \Gamma_2$, act as heat baths at possibly different temperatures $T_1$ and $T_2$ respectively. Particles get absorbed by the heat baths upon collision with $\partial \Gamma$, and, upon collision of a particle with the disk, a new particle is emitted immediately at the collision location with speed $s \in (0,\infty)$ and angle  $\varphi \in (-\frac{\pi}{2},\frac{\pi}{2})$ distributed according to $d(s,\varphi)$ $=\frac{2 \beta_{j(i)}^{3/2}}{\sqrt{\pi}} s^2 e^{-\beta_{j(i)}^2 s^2} \cos(\varphi) ds d\varphi$, where $j(i)=1$ or $2$ depending on whether the particle is confined to the left or the right half of $\Gamma$. This distribution on the boundary $\partial \Gamma$ corresponds to particle's velocity distributed as $\frac{\beta_{j(i)}}{\pi} e^{-\beta_{j(i)}|v|^2} dv$ in $\Gamma$.

We would like to define the associated Markov process with the dynamical rules governed as above. A phase space for such a process should consist of quadruples $(\mathbf{x},\mathbf{v},\theta,\omega)$ with proper identifications at the collisions. In particular, when $x_i \in L_u \cup L_d$, if $v_i$ has positive (negative) horizontal component, then the corresponding particle is confined to the right (left) half of the domain $\Gamma \setminus D$. To simplify the notation, we would like to exclude states $\{(\mathbf{x},\mathbf{v},\theta,\omega):$ for some $i$, $x_i \in L_u \cup L_d$ and $v_i$ has zero horizontal component$\}$ from the phase space. In our future arguments we will frequently omit mentioning the situation of particles reflecting from $L_u \cup L_d$: due to symmetry and circular shape of $\Gamma$, the distance of flight is the same whether reflection occurs or not.

In addition, we would like to exclude all the states with stopped particles ($v_i=0$ for some $i$) and all the states that lead to such with positive probability. A particle stops if it collides tangentially with a stopped disk ($\omega=0$). Consequently, any state $(\mathbf{x},\mathbf{v},\theta,\omega)$ with a particle heading for a tangential collision may lead to positive probability of stopping a particle depending on the particle-disk configuration. For simplicity we exclude all states with particles heading for tangential collisions from the phase space. This ensures that the probability of reaching a state with a stopped particle as system evolves is zero. In addition, we remove zero probability events when particles are moving along the walls $L_u$ and $L_d$. Let
\begin{equa}
\Omega = & \{(\mathbf{x},\mathbf{v},\theta,\omega): v_i \ne 0, 1 \leq i \leq k, \\
& \mathrm{if} \ x_i \in L_u \cup L_d, \ \mathrm{then} \ \mathrm{the} \ \mathrm{horizontal} \ \mathrm{component} \ \mathrm{of} \ v_i \ \mathrm{is} \ \mathrm{nonzero}, \ \mathrm{and} \\
& \mathrm{no} \ \mathrm{particle} \ \mathrm{is} \ \mathrm{heading} \ \mathrm{for} \ \mathrm{a} \ \mathrm{tangential} \  \mathrm{collision} \ \mathrm{with} \ \mathrm{the} \ \mathrm{disk} \} / \sim
\end{equa}
where $\sim$ corresponds to a choice of outgoing velocities upon collision of a particle with $\partial \Gamma \cup \partial D \cup L_u \cup L_d$. The phase space $\Omega$ is forward invariant with probability one.

Let $\Phi_t$ be the associated Markov process on $\Omega$; denote the transition probability kernel by $\mathcal{P}^t$. Note that $\Omega$ is locally compact and separable, and $\Phi_t$ has strong Markov with right-continuous sample paths because we chose to keep track of the outgoing velocities at collisions. Those assumptions are necessary in order to apply general state Markov process theory in section \ref{sect: existence}.

\begin{theorem} \label{thm: main}
There exists a unique absolutely continuous invariant measure $\mu$ for the Markov Process $\Phi_t$. It is mixing, but not exponentially mixing. Moreover, all initial distributions converge to $\mu$, but for a large class of initial distributions the convergence rate is at best polynomial.
\end{theorem}

In section \ref{sect: existence} we will show existence of an invariant probability measure, which is mixing; in section \ref{sect: non exp mixing} that mixing and the convergence of initial distributions to $\mu$ for a class of initial distributions are not exponential. We do not include the proof of absolute continuity of $\mu$ since, given the rest of our results, it is a simple reiteration of the proof of Lemma~12 in \cite{Khanin}.

\section{Existence and Mixing} \label{sect: existence}

\begin{definition}
   A non-empty compact set $C \in \mathcal{B}(\Omega)$, the Borel sigma algebra on $\Omega$, is called $m_C$-petite if there exist $T, \eta>0$ such that
   $$\mathcal{P}^T(x,\cdot) \geq \eta m_C, \;\;\; \forall x \in C.$$
   Here $m_C$ is the uniform probability measure on $C$.
\end{definition}

The condition $\mathcal{P}^T(x,\cdot) \geq \eta m_C$ is frequently called the minorization or Doeblin's condition on $C$.

For any $\delta \geq 0$ and a set $C \in \mathcal{B}(\Omega)$ define $\tau_C$ to be the first hitting time on $C$ and $\tau_C(\delta)$ to be the first hitting time on $C$ after waiting time $\delta$.

We will use the following result by Meyn and Tweedie on continuous-time general state Markov chains, which is a direct consequence of Theorems $1.1$ and $1.2$ in \cite{Meyn_existence}:

\begin{theorem}[\cite{Meyn_existence}] \label{thm: Meyn existence}
Assume there exists a petite set $C$ such that
\begin{itemize}
  \item $\mathbb{P}_{\mathbf{z}}\{\tau_C<\infty\}=1$ for all $\mathbf{z} \in \Omega$ and
  \item for some $\delta>0$, $\sup\limits_{\mathbf{z} \in C}\mathbb{E}_{\mathbf{z}}[\tau_C(\delta)]<\infty$
\end{itemize}
Then there exists an invariant probability measure for $\Phi_t$.
\end{theorem}


We obtain bounds for the expected values of the hitting time of $C$ using the regeneration times idea explained in \ref{subsect: regeneration} and applied rigorously in sections \ref{subsect: tau}, \ref{subsect: Phi_tau returns}, and \ref{subsect: C petite}. Roughly speaking, with the aid of stopping times, we split each random trajectory into similarly behaving pieces of bounded, on average, length. We define the set $C$ and the stopping time $\tau$ and show how regeneration times idea applies in our setting in \ref{subsect: tau}. We estimate the expected times of the lengths of the pieces in \ref{subsect: Phi_tau returns}. In \ref{subsect: C petite} we show that the set $C$ we defined is indeed petite. This ensures the existence of a unique invariant measure $\mu$.

We show that $\mu$ is mixing and all initial distributions converge to it in \ref{subsect: mixing} using a theorem \cite[Thm.~6.1]{MeynII} of Meyn and Tweedie that applies to continuous times Markov processes on general state spaces. To satisfy the conditions on the theorem one needs to demonstrate that some skeleton chain of the Markov process $\Phi_t$ is irreducible. We show that the time-$1$ chain $\Phi_1$ is irreducible using a modification of the argument for showing that $C$ is petite.

\subsection{Regeneration times idea} \label{subsect: regeneration}
To perform the estimates required for Theorem \ref{thm: Meyn existence}, we apply the idea of regeneration times. We will first describe an idealized situation, and then show that a very similar argument works for the real $\Phi_t$. Assume for a moment that there exists a distribution $\nu$ and a stopping time $\tau$ such that $\{\Phi_{t+\tau}, t \geq 0\}$ is independent from $\{\Phi_t, t \geq 0\}$ as well as $\{\Phi_{t+\tau}, t \geq 0\}$ is stochastically equivalent to $\{\Phi_t, t \geq 0\}$ in a sense that they have the same joint distributions provided that $\Phi_0$ is distributed with $\nu$. Then the process can be split into independent regeneration epochs $\{\tau_n\}$. If, in addition,  $\mathbb{E}_{\nu}[\tau] < \infty$ and $\sigma$ is a geometric random variable with $p=\nu(C)$, then
$$\mathbb{E}_{\nu}[\tau_C]=\mathbb{E}_{\nu}[\tau_1 + \cdots + \tau_{\sigma}]$$
$$=\mathbb{P}[\sigma=1]\mathbb{E}_{\nu}[\tau_1] + \mathbb{P}[\sigma=2]\mathbb{E}_{\nu}[\tau_1+\tau_2] + \cdots $$
$$= \nu(C)\mathbb{E}_{\nu}[\tau] + (1-\nu(C))\nu(C)2 \mathbb{E}_{\nu}[\tau] + \cdots = \frac{\mathbb{E}_{\nu}[\tau]}{\nu(C)} <\infty.$$

Theorem \ref{thm: Meyn existence} asks for the initial distribution of $\Phi_0$ to be a point measure at $\mathbf{z} \in \Omega$; assume there exists a stopping time $\tilde{\tau}$ such that $\Phi_{\tilde{\tau}}$ is distributed with $\nu$. If $\sup_{\mathbf{z} \in C} \{\mathbb{E}_{\mathbf{z}}[\tilde{\tau}(\delta)]\}< \infty$ and $\mathbb{E}_{\nu}[\tau] < \infty$, then by a similar estimate we conclude that $\sup_{\mathbf{z} \in C} \{ \mathbb{E}_{\mathbf{z}}[\tau_C(\delta)]<\infty\}$. If, in addition, we show that $\mathbb{P}_{\mathbf{z}}[\tilde{\tau}<\infty]=1$ for all $\mathbf{z} \in \Omega$, then $\mathbb{P}_{\mathbf{z}}\{\tau_C<\infty\}=1$ for all $\mathbf{z} \in \Omega$ follows too.

For the actual process $\Phi_t$, the first part of the argument applies if $\nu$ is the invariant measure. Not only we do not know if it exists, showing that there exists a stopping time $\tilde{\tau}$ with $\Phi_{\tilde{\tau}}$ distributed as $\nu$ is nontrivial. However, if there exists a stopping time $\tau$ such that the system \textquoteleft almost renews' at time $\tau$ since \textquoteleft enough' initial data is forgotten by time $\tau$ due to the randomness of the heat baths, the argument may still carry through. In this situation $\nu$ is not invariant; we rather think of it as a \textquoteleft helper' measure similar to the invariant measure $\mu$.

A bit more precisely we would like to find a stopping time $\tau$ such that $\{\Phi_{t+\tau}, t \geq 0\}$  is \textquoteleft independent enough' from $\{\Phi_t, t \geq 0\}$ and $\Phi_\tau$ is distributed \textquoteleft similar' to $\nu$ given that $\Phi_0$ is distributed \textquoteleft similar' to $\nu$. This will guarantee almost geometric rates of hitting $C$. In addition, we want $\tau$ to be \textquoteleft small enough' so that $\sup_{\mathbf{z} \in C} \{\mathbb{E}_{\mathbf{z}}[\tau(\delta)]\}< \infty$ and $\mathbb{E}_{\nu}[\tau] < \infty$.

\subsection{Proof of existence} \label{subsect: tau}

Let $\mathbf{z}$ be a state in $\Omega$. Some of the particles in $\mathbf{z}$ may be heading for a collision with the disk in a sense that each of these particles will collide with the disk before colliding with $\partial \Gamma$. Let $t_0(\mathbf{z})$ be the time of the last of those collisions with the disk given $\mathbf{z}$. Note that $t_0(\mathbf{z})$ is finite and deterministic.

Let $\tau$ be the minimum time at which all particles and the disk randomize. More precisely, starting with $\mathbf{z}\in \Omega$,
$\tau = \min\{t>0:$ such that both of the following events have occurred:
\begin{itemize}
  \item all particles in $\mathbf{z}$ have collided with $\partial \Gamma$ at least once (ensures that all the initial particles' velocities are forgotten);
  \item a particle originated from $\partial \Gamma$, hit the disk at some time $\tilde{t} > t_0(\mathbf{z})$, and collided with $\partial \Gamma$ again (ensures that the angular velocity $\omega$ is forgotten) \}.
\end{itemize}

A priori it is not completely clear whether $\tau$ is almost surely finite; we will show it along with the expected value estimates. Though, at time $\tau$, the initial velocities of the particles and the angular velocity of the disk are forgotten, the positions $\mathbf{x}$ and $\theta$ may still be strongly correlated since collision times may be. Also note that $\Phi_\tau$ belongs to the collision manifold $\Omega_0 =\{\mathbf{z} \in \Omega: x_i \in \partial \Gamma$ for some $1 \leq i \leq k\}$.

In the following it is convenient to introduce a change of variables in order to make passing from $\Omega$ to $\Omega_0$ easier. We would like to replace $(\mathbf{x},\mathbf{v},\theta,\omega)$ with $(\mathbf{r},\mathbf{s},\mathbf{\varphi},\mathbf{\xi}, \theta, \omega)$, where the new coordinates are based on the information from the past or the future collision. See Fig.\ref{fig: coordinates}. More precisely, let $\mathbf{r}=(r_1, \cdots, r_k)$, $\mathbf{s}=(s_1,\cdots, s_k)$, $\varphi=(\varphi_1, \cdots, \varphi_k)$, and $\xi = (\xi_1, \cdots, \xi_k)$ be as follows: $r_i$ is the point of the past collision of the $i^{th}$ particle with $\partial \Gamma$ if its previous collision was with $\partial \Gamma$ and not with $\partial D$ (here we do not count collisions with $L_u \cup L_d$); otherwise $r_i$ is the point of the future collision of the $i^{th}$ particle with $\partial \Gamma$. Note that the geometry is chosen in such a way that a particle can experience at most one collision with $\partial D$ between collisions with $\partial \Gamma$. In the first scenario, $\xi_i>0$ is the distance of flight of the particle to $\partial \Gamma$ in the direction of $-v_i$ (with possible reflection off $L_u \cup L_d$), and in the second scenario $\xi_i<$ is the distance of flight of the particle to $\partial \Gamma$ in the direction of $v_i$. Let $\varphi \in (-\frac{\pi}{2},\frac{\pi}{2})$ be the angle with respect to the normal to $\partial \Gamma$ at the collision point, and $s_i$ is the speed of the particle. We will denote angles of collision with the disk by $\varphi'$. Note that $\sin(\varphi')=\frac{R+d}{R}\sin(\varphi):= \alpha \sin(\varphi)$.

Then
$$\Omega_0=[\{(\mathbf{r},\mathbf{s},\mathbf{\varphi},\mathbf{\xi},\omega, \theta) \in \Omega: \xi_i=0 \; \mathrm{for} \ \mathrm{some} \ i, \ 1 \leq i \leq k\}. $$

Suppose we start with a point measure $\delta_{\mathbf{z}}$, $\mathbf{z} \in \Omega$, and run $\Phi_t$ until time $\tau$. Once time $\tau$ is reached, the speeds $s_i$ and angles $\phi_i$ of the particles are distributed independently with \\ $d(s_i^{\tau},\varphi_i^{\tau})=\rho_{\beta_{j(i)}}(s_i,\varphi_i)ds_id\varphi_i:=$ $(2 \beta_{j(i)}^{3/2}/\sqrt{\pi}) s_i^2 e^{-\beta_{j(i)}^2 s_i^2} \cos(\varphi_i) ds_i d\varphi_i$ if $\xi_i > 0$ and $|\omega|$ is distributed with  $d|\omega^{\tau}|=\rho_{\beta_j}(\omega)d\omega=$ $(2\sqrt{\beta_j}/\sqrt{\pi})e^{-\beta_j \omega^2}d\omega$, where $j=1$ or $2$ depending on the side from which the disk experienced its last collision before time $\tau$ (in case of $\xi_i<0$, i.e. after disk collision, $d(s_i^{\tau},\varphi_i^{\tau})$ is the same if $\beta_{j(i)} = \beta_j$  and is a certain mix distribution if $\beta_{j(i)} \ne \beta_j$; it is similar to $\rho_{\beta_{j(i)}}(s_i,\varphi_i)ds_id\varphi_i$ and can be bounded above and below their common upper and lower bounds, but cannot be written in a closed form). The inverse temperature $\beta_j$ is the only memory kept for the distribution of $|\omega|$ at time $\tau$.

\medbreak

Let $\mu_{\mathbf{z}}:=\mathcal{P}^{\tau}_* \delta_{\mathbf{z}}$. We would like to show that due to \textquoteleft enough randomization' of speeds and angular velocities, the expected time $\mathbb{E}_{\mu_{\mathbf{z}}}[\tau]$ has a uniform upper bound for all initial $\mathbf{z}$. The same holds for all subsequent $\tau$-stops. More generally,

\begin{lemma} \label{lemma: subsequent tau}
  There exists $D>0$ such that for any initial distribution $\lambda$, if $\nu = \mathcal{P}^{\tau}_* \lambda$, then $\mathbb{E}_{\nu}[\tau]<D$.
\end{lemma}

Note that Lemma \ref{lemma: subsequent tau} does not say anything about the initial waiting time for renewal $\tau$ starting from $\mathbf{z} \in \Omega$ or an arbitrary initial distribution $\lambda$, only about the subsequent ones. We will prove Lemma \ref{lemma: subsequent tau} in subsection \ref{subsect: Phi_tau returns}.

In order to apply Theorem \ref{thm: Meyn existence}, we need to find a petite set $C$ such that probabilities of hitting $C$ at regeneration times $\tau_n$ are uniformly bounded away from $0$ and $1$. In addition, we want $\sup_{\mathbf{z} \in C}\{\mathbb{E}_{\mathbf{z}}[\tau(\delta)]\} < \infty$. Roughly speaking, we expect $C$ to be a collection of all states $\mathbf{z}$ in $\Omega$ with uniform upper and lower bounds on particles' speeds and an upper bound on disk's angular velocity both before and after collision. Since $\Phi_\tau$ has values on $\Omega_0$, but, in reality, we are concerned with $C$ defined on whole $\Omega$, it would be convenient if $C$ was forward invariant during particle's flight from $\partial \Gamma$ to $\partial \Gamma$.

If a particle collides with a nearly stopped disk, $\omega \approx 0$, nearly tangentially, $\varphi' \approx \pm \frac{\pi}{2}$, then the particle's velocity is significantly reduced, $s' \approx 0$, and it would take a very long time for such a particle to reach $\partial \Gamma$. So when defining $C$, we would like to restrict disk collisions from being too close to tangential. In fact, we would like $|\sin(\varphi)|$ to be bounded away from $\frac{\alpha}{\sqrt{1+\alpha^2}}$, where $\alpha=\frac{R}{R+d}$; this will be used later in the proof.

\medbreak

Let $C$ to be the set of all $\mathbf{z} \in \Omega$ such that for some $s_{\min}<s_{\max}$ and $\epsilon>0$, $|\omega| \leq s_{\max}\sqrt{\frac{1-\epsilon}{1+\alpha^2}}$ and for all $i$, $1 \leq i \leq k$:

\begin{itemize}
  \item either $|\sin(\varphi)| > \alpha$ (no collision with the disk) and $s_{\min} \leq s_i \leq s_{\max}$,
  \item or $|\sin(\varphi)| < \frac{\alpha}{\sqrt{1+\alpha^2}} \sqrt{1-\epsilon}$, $\xi_i>0$ (before collision with the disk), and $s_{\min} \leq s_i \leq s_{\max}$,
  \item or $|\sin(\varphi)| < \frac{\alpha \sqrt{1-\epsilon}}{(1+\alpha^2)\sqrt{s_{\max}^2 (1-\epsilon)+s_{\min}^2(\alpha^2 + \epsilon)}}$, $\xi_i<0$ (after collision with the disk), and $s_{\min} \sqrt{\frac{\alpha^2+\epsilon}{\alpha^2+1}} \leq s_i \leq s_{\max} \sqrt{\frac{\alpha^2-\epsilon+2}{\alpha^2+1}}$.
\end{itemize}

\begin{proposition} \label{prop: petite}
  $C$ is petite.
\end{proposition}

We will prove Prop. \ref{prop: petite} in subsection \ref{subsect: C petite}.

Easy computation ensures that $C$ is forward invariant between collisions of particles with $\partial \Gamma$. Moreover, the only way to leave $C$ is for some particle to collide with $\partial \Gamma$ and to originate with new speed and angle not satisfying the speed and angle conditions above.

Similarly, in order for $\Phi_\tau \in C$, all the particles' speeds and angles must be drawn in the correct range. Since the drawings are independent, there are a lower bound $\gamma_{\min}$ and an upper bound $\gamma_{\max}$ on the probability for $\Phi_{\tau} \in C$.

To show that the conditions of Theorem \ref{thm: Meyn existence} are satisfied, we need to estimate the expected value of the first hitting time $\tau$ provided we originate in $C$ and wait some small time $\delta$.

\begin{lemma} \label{lemma: tau(delta) from C}
 There exists $\delta>0$ and $D'>0$ such that
 $$\sup\limits_{\mathbf{z} \in C}\{\mathbb{E}_{\mathbf{z}}[\tau(\delta)]\} \leq D'.$$
Moreover, $\mathbb{P}_{\mathbf{z}}\{\tau(\delta) <\infty\}=1$.
\end{lemma}

We will prove Lemma \ref{lemma: tau(delta) from C} in subsection \ref{subsect: Phi_tau returns}.

\medbreak

Let $\sigma_C$ be the hitting time of the set $C$ for the stopped Markov chain $\Phi_\tau$, i.e. $\sigma_C=\min\{n >0: (\Phi_{\tau})^n \in C\}$.
Then for $\mathbf{z} \in C$,

$$\mathbb{E}_{z}[\tau_C(\delta)] \leq \mathbb{P}(\sigma_C=1)\mathbb{E}_{\mathbf{z}}[\tau(\delta)] + \mathbb{P}(\sigma_C=2)(\mathbb{E}_{\mathbf{z}}[\tau(\delta)]+\mathbb{E}_{\mathcal{P}_*^{\tau}\delta_{\mathbf{z}}}[\tau])$$
$$+ \mathbb{P}(\sigma_C=3)(\mathbb{E}_{\mathbf{z}}[\tau(\delta)]+\mathbb{E}_{\mathcal{P}_*^{\tau}\delta_{\mathbf{z}}}[\tau]+\mathbb{E}_{\mathcal{P}_*^{2\tau}\delta_{\mathbf{z}}}[\tau])+ \cdots $$
$$\leq \gamma_{\max} D' + (1-\gamma_{\min})\gamma_{\max}(D'+D)+ (1-\gamma_{\min})^2 \gamma_{\max} (D'+2D)  +\cdots $$
$$= \gamma_{\max} D'[1+(1-\gamma_{\min})+(1-\gamma_{\min})^2 \cdots]+ (1-\gamma_{\min})\gamma_{\max} D[1+2(1-\gamma_{\min})+ \cdots]$$
$$=\frac{\gamma_{\max}}{\gamma_{\min}}D' + \frac{(1-\gamma_{\min})\gamma_{\max}}{\gamma_{\min}^2}D.$$

This is the second condition of the Theorem \ref{thm: Meyn existence}. The first condition follows from $\mathbb{P}_{\mathbf{z}}\{\tau(\delta) <\infty\}=1$ in Lemma \ref{lemma: tau(delta) from C} and the estimates above. We conclude that there exists an invariant probability measure for $\Phi_t$. $\square$

\subsection{Estimates for $\mathbb{E}_{\mathbf{z}}(\tau)$} \label{subsect: Phi_tau returns}

Before proceeding with estimating $\mathbb{E}_{\mathbf{z}}(\tau)$, let us first estimate the expected value of a flight time from $\partial \Gamma$ to $\partial \Gamma$ by a particle emitted randomly. Let $T^{\mathrm{flight}}$ be the random stopping time of hitting $\partial \Gamma$ by a particle emitted at time $0$ with velocity distribution $d(s,\varphi)=\rho_{\beta_j}(s,\varphi)ds d\varphi$ that will hit the disk rotating at angular velocity $\omega$. Let $l:=\sqrt{(R+d)^2-R^2}$ be half of the maximal distance of flight. Then

\begin{equa}
\mathbb{E}[T^{\mathrm{flight}}] & \leq  \int\limits_{\sin(\varphi)\geq \alpha} \int_{0}^{\infty} \frac{2l}{s} \rho_{\beta_j}(s,\varphi)ds d\varphi \\
& + \int\limits_{\sin(\varphi)\leq \alpha} \int_{0}^{\infty} [\frac{l}{s}+\frac{l}{\sqrt{\omega^2 + s^2\cos^2(\varphi')}}]\rho_{\beta_j}(s,\varphi)ds d\varphi. \\
\end{equa}

Note that $$\frac{l}{\sqrt{\omega^2 + s^2\cos^2(\varphi')}} \leq \frac{l}{ s \cos(\varphi')}=\frac{l}{ s \sqrt{1-\sin^2(\varphi')}}=\frac{l}{ s \sqrt{1-\sin^2(\varphi)/\alpha^2}}.$$

Therefore

$$\mathbb{E}[T^{\mathrm{flight}}] \leq \int\limits_{\sin(\varphi)\geq \alpha} \int_{0}^{\infty} \frac{2l}{s} \frac{2 \beta_j^{3/2}}{\sqrt{\pi}}s^2 e^{-\beta_j s^2}\cos(\varphi) d \varphi ds $$
$$+ \int\limits_{\sin(\varphi)\leq \alpha} \int_{0}^{\infty} [\frac{l}{s}+\frac{l}{ s \sqrt{1-\sin^2(\varphi)/\alpha^2}}]\frac{2 \beta_j^{3/2}}{\sqrt{\pi}}s^2 e^{-\beta_j s^2}\cos(\varphi) d \varphi ds \leq K,$$
for some $K>0$ independent on $\omega$.

This guarantees, in particular, that every randomly emitted particle has finite expected time of flight.

Let $T^{\mathrm{hit}}$ be a random stopping time of hitting $\partial \Gamma$ after one disk collision by a particle emitted from $\partial \Gamma$ with $(s,\varphi) \sim \rho_{\beta_j}(s,\varphi)ds d\varphi$. The probability of hitting the disk at each round is $\alpha$, so

$$\mathbb{E}[T^{\mathrm{hit}}] \leq \alpha K + (1-\alpha)\alpha 2K + (1-\alpha)^2 \alpha 3K + \cdots $$
$$= \alpha K[1 + 2(1-\alpha)+3(1-\alpha^2)+ \cdots]=\frac{\alpha K}{\alpha^2}=\frac{K}{\alpha}.$$

\medbreak

Now we are ready for preliminary estimates for $\mathbb{E}_{\mathbf{z}}(\tau)$ for any $\mathbf{z} \in \Omega$.

First note that $\tau \leq \max_{1 \leq i \leq k}\{T^i\}+T^{\mathrm{flight}}+T^{\mathrm{hit}}$, where $T^i$ is the random time it takes for particle $i$ in $\mathbf{z}$ to reach $\partial \Gamma$. Indeed, if no random particle collided with the disk before the last one hit $\partial \Gamma$, then either the last particle will hit the disk in time $T^{\mathrm{hit}}$ or some other one will before; it will take less than $T^{\mathrm{flight}}$ time for it to exit. Therefore

$$\mathbb{E}_{\mathbf{z}}[\tau] \leq \max\limits_{1 \leq i \leq k}\{\mathbb{E}_{\mathbf{z}}[T^i]\}+\mathbb{E}[T^{\mathrm{flight}}]+\mathbb{E}[T^{\mathrm{hit}}]$$

\medbreak

Let us estimate $\mathbb{E}_{\mathbf{z}}[T^1]$. The initial position and velocity of the particle uniquely determine whether the particle will collide with the disk or not. In case of no collision, the time of flight is $ \leq \frac{2l}{s_1}$. If the particle is headed for a collision with the disk, there are three possibilities for the value of the angular velocity of the disk upon collision:
\begin{enumerate}
  \item original $\omega$;
  \item $s_2 \sin(\varphi_2')$, if a particle, which we label as $2^{nd}$, collided with the disk immediately before the first particle; and
  \item a random angular velocity $\tilde{\omega}$ acquired by the disk due to collision with a particle emitted randomly from $\partial \Gamma$.
\end{enumerate}

In the $3^{rd}$ situation, $|\tilde{\omega}|$ is drawn from the distribution $2\sqrt{\frac{\beta_j}{\pi}} e^{-\beta_j \tilde{\omega}^2}d\tilde{\omega}$. The expected value for the exit time after collision with the disk is bounded as follows:
$$\int_{-\infty}^{\infty}\frac{l}{\sqrt{\tilde{\omega}^2+s_1^2\cos^2(\varphi_1')}} \sqrt{\frac{\beta_j}{\pi}} e^{-\beta_j \tilde{\omega}^2}d\tilde{\omega} $$
$$\leq 2\int_{0}^{\infty}\frac{l}{\tilde{\omega}} \sqrt{\frac{\beta_j}{\pi}} e^{-\beta_j \tilde{\omega}^2}d\tilde{\omega} \leq \frac{l}{\sqrt{\beta_j \pi}} \leq \frac{l}{\sqrt{\beta_{\min} \pi}}$$

For the convenience of notation below, set $\varphi' \equiv 0$ if $\sin(\varphi) \geq \frac{R}{R+d}$. Then

$$\mathbb{E}_{\mathbf{z}}[T^1] \leq \frac{l}{s_1}+  \max\{\frac{l}{s_1}, \frac{l}{\sqrt{\omega^2+s_1^2\cos^2(\varphi_1')}}, \frac{l}{\sqrt{s_2^2 \sin^2(\varphi_2') +s_1^2\cos^2(\varphi_1')}},\frac{l}{\sqrt{\beta_{\min} \pi}}\} $$
$$\leq \frac{2l}{s_1 \cos(\varphi_1')}+\frac{l}{\sqrt{\beta_{\min} \pi}}$$
Similar estimate holds for $\mathbb{E}_{\mathbf{z}}[T^i]$, $2 \leq i \leq k$.

Therefore
\begin{equation} \label{eqn: E tau bound}
\mathbb{E}_{\mathbf{z}}[\tau] \leq \max\limits_{1 \leq i \leq k} \{\frac{2l}{s_i \cos(\varphi_i')}\}+\frac{l}{\sqrt{\beta_{\min} \pi}}+\frac{(1+\alpha)K}{\alpha}
\end{equation}

\medbreak

\textbf{Proof of Lemma \ref{lemma: subsequent tau}.
}

The bound $\mathbb{E}_{\mathbf{z}}[\tau]$ in the equation (\ref{eqn: E tau bound}) depends only on $s_i$ and  $\cos(\varphi_i')$, $1 \leq i \leq k$. In the distribution $\nu = \mathcal{P}^{\tau}_* \lambda$, each $s_i$ and $\varphi_i$ are distributed according to $\rho_{\beta_j(i)}(s_i, \varphi_i)ds_i d\varphi_i$ (before disk collision).

Then

\begin{equa}
\mathbb{E}_{\nu}[\tau] & \leq \int\limits_{\Omega} [\max\limits_{1 \leq i \leq k} \{\frac{2l}{s_i \cos(\varphi_i')}\} + \frac{l}{\sqrt{\beta_{\min} \pi}}+\frac{(1+\alpha)K}{\alpha}]d\nu \\
& \leq \int_{-\frac{\pi}{2}}^{\frac{\pi}{2}} \int_{0}^{\infty} \frac{2kl}{s \cos(\varphi')} \rho_{\beta_j}(s,\varphi)dr d\varphi+\frac{l}{\sqrt{\beta_{\min} \pi}}+\frac{(1+\alpha)K}{\alpha} \\
& = \int_{0}^{\arcsin{\alpha}} \int_{0}^{\infty} \frac{2kl}{s \sqrt{1-\sin^2(\varphi)/\alpha^2}}\frac{2 \beta_j^{3/2}}{\sqrt{\pi}}s^2 e^{-\beta_j s^2} \cos(\varphi) d\varphi ds \\
& +
 \int_{\arcsin{\alpha}}^{\frac{\pi}{2}} \int_{0}^{\infty} \frac{2kl}{s}\frac{2 \beta_j^{3/2}}{\sqrt{\pi}}s^2 e^{-\beta_j s^2} \cos(\varphi) d\varphi ds +\frac{l}{\sqrt{\beta_{\min} \pi}}+\frac{(1+\alpha)K}{\alpha} \leq D
\end{equa}
for some $D>0$. $\square$

\medbreak

\textbf{Proof of Lemma \ref{lemma: tau(delta) from C}.}

If $\mathbf{z} \in C$ then
\begin{equa} \label{eqn: E tau bound C}
\mathbb{E}_{\mathbf{z}}[\tau] & \leq \max\limits_{1 \leq i \leq k} \{\frac{l}{s_i \cos(\varphi_i')}\}+\frac{2l}{\sqrt{\beta_{\min} \pi}}+\frac{(1+\alpha)K}{\alpha} \\
& \leq \frac{2l}{\sqrt{\epsilon} s_{\min}}+\frac{l}{\sqrt{\beta_{\min} \pi}}+\frac{(1+\alpha)K}{\alpha}
\end{equa}

If we start with $\mathbf{z} \in C$ and wait for some time $\delta$, then some of the particles may experience collisions with $\partial \Gamma$ and redistribute their speeds and angles according to $\rho_{\beta_j(i)}(s,\varphi)dr d\varphi$.
Then
$$\mathbb{E}_{\mathbf{z}}[\tau(\delta)] \leq \frac{2l}{\sqrt{\epsilon}s_{\min}}+D =: D',$$

where $D$ is the constant from Lemma \ref{lemma: subsequent tau}.

Now suppose we start with any $\mathbf{z} \in \Omega$. The stopping time satisfies $\tau \leq \max_{1 \leq i \leq k}\{T^i\}+T^{\mathrm{flight}}+T^{\mathrm{hit}}$. The time $T^i \leq \frac{2l}{s_i \cos(\varphi_i')}+\frac{l}{\tilde{\omega}}$ is finite almost surely; so are $\mathbb{E}[T^{\mathrm{flight}}]$ and $\mathbb{E}[T^{\mathrm{flight}}]$ since their expectations are bounded. For almost every $\mathbf{y}$ in the support of $\mathcal{P}^{\delta}(\mathbf{z},\cdot)$, $\tau$ is also finite almost surely. Therefore, for any $\mathbf{z} \in \Omega$, $\mathbb{P}_{\mathbf{z}}\{\tau(\delta) <\infty\}=1$. $\square$

\medbreak

\subsection{C is petite: proof of Prop. \ref{prop: petite}} \label{subsect: C petite}
We would like to show that there exits $T, \eta>0$ such that for any $\mathbf{z} \in C$, $\mathcal{P}^T(\mathbf{z},\cdot) \geq \eta m_C,$ where $m_C$ is the uniform probability measure on $C$. This statement implies, in particular, that for any $\mathbf{z}, \mathbf{z}' \in C$, there is a sample path $\sigma(\mathbf{z},\mathbf{z}',T)$ which takes precisely time $T$ to complete. We will start by showing this implication, with additional restrictions for the path to be \textquoteleft regular' in a sense that it stays away from tangential collisions (precise definition to follow) and for $\sigma(\mathbf{z},\mathbf{z}',T) \subset C$. Since $\eta$ can be chosen to be arbitrarily small, we can to restrict our view of the dynamics along $\sigma(\mathbf{z},\mathbf{z}',T) \subset C$ and ignore the trajectories that diverge from $\sigma(\mathbf{z},\mathbf{z}',T) \subset C$.  In this sense, our proof is of a similar flavor as the proof of the minorization condition, Prop.~2, in \cite{Khanin}.

The state $\mathbf{z} \in C$ represents positions and velocities of $k$ particles as well as the marked position and the angular velocity of the disk. Particles and the disk interact as the dynamics evolves. In order to make the analysis simpler, we would like to choose a path that \textquoteleft decouples' the particles and the disk. We achieve this by imposing a rule that as soon as a particle reaches $\partial \Gamma$, it does not collide with the disk anymore. With two exceptions. One is that some particle needs to reset the disk's angular velocity and the other is that some particles need to collide with the disk after their last collisions with $\partial \Gamma$ in order to reach $\mathbf{z}'$. Note that, with the above assumption, if $\mathbf{z} \in C$, the times $t_i$ for $i^{th}$ particle to reach $\partial \Gamma$ for the first time are deterministic and uniformly bounded by some $\tilde{t}>0$. We treat the final state $\mathbf{z}'$ similarly by running the dynamics backwards in time. Let $t_i'$ be the times for the particles in $\mathbf{z}'$ to reach $\partial \Gamma$ when running the system backwards in time; note that $t_0' \leq \tilde{t}$.

In addition, we impose that, at each collision with $\partial \Gamma$, with the exception of setting the disk's angular velocity and the last collision for each particle, the emission angles are uniformly bounded away from $\frac{R}{R+d}$ and $1$. Namely, we require that the emission angles satisfy $\alpha + \epsilon \leq \sin(\varphi) \leq 1-\epsilon$ for $\alpha=\frac{R}{R+d}$ and some $\epsilon>0$.

The \textquoteleft decoupling' of the particles reduces the problem to a sub-problem concerning only one particle, which can be stated as follows:

\begin{lemma} \label{lemma: path subproblem}
For large enough $\tilde{T}$ and for any $r, r' \in \partial \Gamma_L $ (or $\partial \Gamma_R$), there exists a particle path $\sigma_i$ from $r$ to $r'$, with outgoing angles and speeds satisfying $\alpha + \epsilon \leq |\sin(\varphi)| \leq 1-\epsilon$ and $s_{\min} \leq s \leq s_{\max}$, such that $\sigma_i$ takes precisely time $\tilde{T}$ to complete. Moreover, the paths can be chosen in such a way that the number of collisions is bounded by some monotone function of $\tilde{T}$.
\end{lemma}

\textbf{Proof of Lemma \ref{lemma: path subproblem}.}

Let $m$ be the minimal number of collisions required to travel between two diametrically opposite points satisfying the angle assumption $\alpha + \epsilon \leq |\sin(\varphi)| \leq 1-\epsilon$ (usually $m=2$, but if $d$ is very small compared to $R$, $m$ may be larger). Then $m$ collisions is enough to travel between any $r$ and $r'$. Let $t_{\min}$ and $t_{\max}$ be the minimal and maximal times of travel along such a path with $\leq m$ collisions satisfying the speed and angle parameters of Lemma \ref{lemma: path subproblem}. Note that a simple application of the Lagrange multipliers method guarantees that, for all intermediate values $t \in [t_{\min}, t_{\max}]$, there exists a path taking precisely time $t$.

Let $\overline{t}_{\min}$ and $\overline{t}_{\max}$ be the minimal and maximal times of travel along a back and forth path originating and ending at $r'$, satisfying $\alpha + \epsilon \leq |\sin(\varphi)| \leq 1-\epsilon$ ($\overline{t}_{\min}$ and $\overline{t}_{\max}$ do not depend on the choice of $r'$). By appending this path to end of the path between $r$ and $r'$, we can produce paths from $r$ to $r'$ taking any $t \in [t_{\min}+k\overline{t}_{\min},t_{\max}+k\overline{t}_{\max}]$ with $m+2k$ segments. Let $\overline{k} = \min\{k: \overline{t}_{\min} < (t_{\max}-t_{\min})+ k(\overline{t}_{\max}-\overline{t}_{\min})\}$. Then for any $\tilde{T} \geq t_{\min}+\overline{k}\overline{t}_{\min}$, there exists a path between $r$ and $r'$ taking time $\tilde{T}$ to complete with the maximal number of segments $m+2\lfloor \frac{T-t_{\min}}{\overline{t}_{\min}} \rfloor$, as desired. $\square$

\medbreak

Given Lemma \ref{lemma: path subproblem}, in order to reach $\mathbf{z}'$, we just need to reconnect the paths of particles $1$ through $k$  originating at $\mathbf{z}$ to the paths of particles $1$ through $k$ ending at $\mathbf{z}'$. In addition, we need to select one particle $p_{\omega}$, which will be sent to reset the disk to the desired angular velocity $\omega'$; let $p_{\omega}$ be the particle with the latest first collision with $\partial \Gamma$. For particles that do not reset the disk, let $\tilde{T} = T-t_i-t_i'$.

The particle $p_{\omega}$ that resets the disk has to change the disk's angular velocity from $\omega$ to $\omega'$, where $\omega$ and $\omega'$ are the angular velocities of the disk after all the particles in $\mathbf{z}$ or $\mathbf{z}'$ have collided with $\partial \Gamma$ in forward and backward times respectively. To change $\omega$ to $\omega'$ one simply needs to emit a particle with $(r,\xi,s,\varphi) \in C$ such that $\omega'=s \sin(\varphi')=s \sin(\varphi)/\alpha$.

However, we are also required to keep track of the changes in the disk's angular position $\theta$ so that it to matches when we reach the final state $\mathbf{z}'$. Additional challenges arise from estimating the densities along $\sigma(\mathbf{z},\mathbf{z}',T) \subset C$ and require dealing with bounds on Jacobians of functions of many variables. In order to simplify our arguments later, we choose to first reset $\omega$ to some intermediate value $\tilde{\omega}$ and then send $p_{\omega}$ again to reset from $\tilde{\omega}$ to $\omega'$. We require that the outgoing angles for resetting the angular velocities satisfy $|\sin(\phi)| \leq \frac{\alpha \sqrt{1-\epsilon}}{\sqrt{1+\alpha^2}}$ and, for some $\kappa>0$ (provided in Lemma \ref{lemma: reacquiring density}), $\tilde{\omega} \in [-2\kappa, 2\kappa]$, $|\omega - \tilde{\omega}|>\kappa$ and $|\omega' - \tilde{\omega}|>\kappa$.

The path of the particle $p_{\omega}$ is as follows: after its first collision with $\partial \Gamma$, $p_{\omega}$ originates with $\varphi$, $\alpha + \epsilon \leq |\sin(\varphi)| \leq 1 - \epsilon$, and collides with $\partial \Gamma$ again at time $t_{p_{\omega}}$ (at this collision $p_{\omega}$ \textquoteleft acquires density'; see Lemma \ref{lemma: acquiring density}). Then $p_{\omega}$ is sent to reset the angular velocity of the disk first to $\tilde{\omega}$ and then to $\omega'$ in total time $t^{\mathrm{set}}$. Let $\Delta \theta^{\mathrm{reset}}$ be the change in the angular position of the disk during the pure angular velocity resetting. In addition, we let $p_{\omega}$ to wait time  $t^{\mathrm{wait}}$ between resetting to $\tilde{\omega}$ and resetting to $\omega'$. The waiting time $t^{\mathrm{wait}}$ is helpful for matching the angular position of the disk $\theta'$ at time $T$ and the while \textquoteleft waiting', $p_{\omega}$ essentially flies back and forth hitting $\partial \Gamma$ as in the proof of Lemma \ref{lemma: path subproblem}. After $p_{\omega}$ resets the disk, it follows a path from Lemma \ref{lemma: path subproblem} and then flies to the position from $\mathbf{z}'$ assigned to $p_{\omega}$. Then,

$$\theta' + 2\pi n = \theta + t^{\mathrm{wait}} \tilde{\omega} + \Delta \theta^{\mathrm{reset}} + (T-t_{p_{\omega}} - t^{\mathrm{set}} - t^{\mathrm{wait}})\omega'$$

This sets $t^{\mathrm{wait}} = \frac{\mathrm{const}}{\tilde{\omega} - \omega'} \leq \frac{2 \pi}{\kappa}$.

Note that $t^{\mathrm{set}}_{\min} \leq t^{\mathrm{set}} \leq t^{\mathrm{set}}_{\max}$, for some $t^{\mathrm{set}}_{\min}, t^{\mathrm{set}}_{\max}>0$. Let $\tilde{T} = T-t_{p_{\omega}}-t_{p_{\omega}'}-t^{\mathrm{set}}$ for $p_{\omega}$.

If we take $T>2\tilde{t}+t^{\mathrm{set}}_{\max}+ \frac{2\pi}{\kappa}+ t_{\min}+\overline{k}\overline{t}_{\min}$, then we are guaranteed to have a path from $\mathbf{z}$ to $\mathbf{z}'$ with $\leq N := m+2\lfloor \frac{T-t_{\min}+ t^{\mathrm{wait}}}{\overline{t}_{\min}} \rfloor + 3$ collisions with $\partial \Gamma$. Call such a path $\sigma(\mathbf{z},\mathbf{z}',T)$.

Thus, we obtain:

\begin{lemma} \label{lemma: path}
Given $\epsilon>0$, there exists $T>0$ and $N>0$ such that for any $\mathbf{z},\mathbf{z}' \in C$, there exists a sample path $\sigma(\mathbf{z},\mathbf{z}', T)$ making less than $N$ collisions such that between the first and the last collisions of each particle with the $\partial \Gamma$, all $s_{\min} \leq s_i \leq s_{\max}$ and $\alpha + \epsilon \leq |\sin(\varphi)| \leq 1-\epsilon$.
\end{lemma}

\medbreak

\textbf{Density bounds along $\sigma(\mathbf{z},\mathbf{z}', T)$.}

The next step is to show that if we start with a point measure at $\mathbf{z}$, $\delta_{\mathbf{z}}$, it \textquoteleft acquires density' as it evolves with the dynamics in a sense that $\mathcal{P}_*^t \delta_{\mathbf{z}}$ has a nontrivial absolutely continuous component for large enough $t$. Moreover, the density of this absolutely continuous component is uniformly bounded away from zero in a neighborhood of each path $\sigma(\mathbf{z},\mathbf{z}',T)$ and, in particular, at the endpoint $\mathbf{z}' \in C$.

Density in a neighborhood of a point $\tilde{\mathbf{z}} \in \Omega$ is the product of the densities in neighborhoods of coordinates of each particle and the disk. For the majority of the path $\sigma(\mathbf{z},\mathbf{z}',T)$, particles and the disk do not interact, so we can deal with their densities separately. In Lemma \ref{lemma: acquiring density} we show that each particle acquires density with a uniform lower bound in a fixed size neighborhood of the second collision and Lemma \ref{lemma: pushing density forward} keeps track of lower bounds of the densities at subsequent collisions along the path as we push the measure forward. Lemmas \ref{lemma: acquiring density for the disk} and \ref{lemma: reacquiring density} deal with the particle $p_{\omega}$ that collides with the disk, making the disk acquire density and the particle to re-acquire density at the next collision with $\partial \Gamma$ after loosing some to the disk. Since the number of collisions along $\sigma(\mathbf{z},\mathbf{z}',T)$ is bounded, if we combine Lemmas \ref{lemma: acquiring density}-\ref{lemma: reacquiring density} together, by the last collision, each particle as well as the disk have density in a fixed-size neighborhood of the collision point. After the last collision the dynamics is deterministic and the value of the density is preserved under push forwards. This allows us to conclude that $\mathcal{P}_*^T \delta_{\mathbf{z}}$ has a uniform lower bound on the density at $\mathbf{z}'$.

To formalize the argument, we need to define more precisely what we mean by a neighborhood of a collision point: in the coordinate system defined in section \ref{sect: settings} the dynamics has discontinuities at collision points. Let us first replace $\varphi$ coordinate by $\sin(\varphi)$: this coordinate change makes $\sin(\varphi)$ re-distribute uniformly at collisions; in addition, for $(r,\sin(\varphi),\xi)$-coordinates, the Jacobian for the standard billiard flow is equal to $1$ (see \cite{Chernov} for details). Then, at each collision point of $\sigma(\mathbf{z},\mathbf{z}',T)$, we extend the coordinates $(r,\xi,s,\sin(\varphi))$ forward or backward in time to accommodate neighborhoods of fixed size $\zeta$ for some small enough $\zeta>0$. Here we assume that $\xi$ is taking negative values before collisions and positive values after collisions. This extension is possible due to the bounds on the speeds and angles introduced in Lemma \ref{lemma: path}.

First, we show that by pushing $\delta_{\mathbf{z}}$ forward, one can acquire density with a uniform lower bound. By the design of our path $\sigma(\mathbf{z},\mathbf{z}',T)$, as soon as a particle reaches $\partial \Gamma$, it is independent from other particles. Therefore, we only need to show that each particle acquires density with a uniform lower bound in a $\zeta$-neighborhood of some point along a particle sub-path in $\sigma(\mathbf{z},\mathbf{z}',T)$.

Given $\mathbf{z}=(r,0,s,\sin(\varphi)) \in \Omega_0$ and $\zeta>0$, let
$$H_{\zeta}(r,0,s,\sin(\varphi)):=(r-\zeta,r+\zeta) \times (-\zeta,+\zeta) \times (s-\zeta,s+\zeta)\times (\sin(\varphi)-\zeta,\sin(\varphi)+\zeta)$$
and let $\mu_{H_{\zeta}(r,0,s,\sin(\varphi))}$ be the uniform measure with density $1$ on $H$.

\begin{lemma} \label{lemma: acquiring density}
There exist $\zeta_0, \eta_0$ such that $\Pi_i[\mathcal{P}_*^{\tau_i} \delta_{\mathbf{z}}] \geq \eta_0 \mu_{H_{\zeta_0}(r,0,s,\sin(\varphi))}$, $1 \leq i \leq k$, where $\tau_i$ is the time of the second collision of particle $i$ with $\partial \Gamma$ along $\sigma(\mathbf{z},\mathbf{z}',T)$, $(r,0,s,\sin(\varphi))$ - the coordinates of the collision, and $\Pi_i$ is the projection to the coordinates of the $i^{th}$ particle.
\end{lemma}

We will prove Lemma \ref{lemma: acquiring density} as well as all other technical lemmas in this argument at the end of the subsection.

In the following, we will slightly abuse the notation by using the operator $\mathcal{P}_*$ to push forward measures projected on components of $\Omega$ associated with a single particle or a particle and the disk. This operation is well defined since we will only do so on time intervals for which the chosen component does not interact with the rest of the system.

In Lemma \ref{lemma: pushing density forward} we show that if we start with a uniform measure $\nu$ with density $1$ in a fixed-size neighborhood of a collision point of $\sigma(\mathbf{z},\mathbf{z}',T)$, then its push forward at the next collision point  has a uniform lower bound on the density in some uniformly sized neighborhood.

\begin{lemma} \label{lemma: pushing density forward}
Let $l$ be a line segment connecting two points $r$ and $r'$ in $\partial \Gamma$ forming an angle $\varphi$ with $\partial \Gamma$, satisfying $\alpha + \epsilon \leq |\sin(\varphi)| \leq 1-\epsilon$. Let $s$, $s_{\min} \leq s \leq s_{\max}$, be the speed drawn at $r$ and let $t$ be the time to trace the segment $l$ with speed $s$. Let $s'$, $s_{\min} \leq s' \leq s_{\max}$, and $\varphi'$, $\alpha + \epsilon \leq |\sin(\varphi')| \leq 1-\epsilon$, be the new speed and angle drawn at $r'$.

Then for any $\zeta>0$ with $H_{\zeta}(r,0,s,\sin(\varphi))$ well defined, there exist $\eta', \zeta'>0$ such that $\mathcal{P}_*^t \mu_{H_{\zeta}(r,0,s,\sin(\varphi))} \geq \eta' \mu_{H_{\zeta'}(r',0,s',\sin(\varphi'))}$.
\end{lemma}

Since Lemma \ref{lemma: path} guarantees a uniform upper bound $N$ on the number of collisions with $\partial \Gamma$ along $\sigma(\mathbf{z}, \mathbf{z}', T)$, Lemmas \ref{lemma: acquiring density} and \ref{lemma: pushing density forward} guarantee uniform lower bounds on the densities along $\sigma(\mathbf{z}, \mathbf{z}', T)$ up until last particle collisions.  It remains to treat the special case of the particle $p_{\omega}$ that resets the disk's angular velocity.

Let $r_0$ be the point of the second collision with $\partial \Gamma$ of $p_{\omega}$. Let $s_0$ and $\sin(\varphi_0)$ be parameters required for the particle to reset the disk to $\tilde{\omega}$. Let $d_1 \in \partial D$ be the location of the disk collision, $t_0$ be the time of flight between $r_0$ and $d_1$, $r_2  \in \partial \Gamma$ be the location of the following collision with $\partial \Gamma$, $t_1$ the time of flight from $d_1$ to $r_2$, and $\tilde{\theta}$ be the position of the marked point on the disk at time $t_0+t_1$. Then the outgoing speed at $d_1$ is $s_1=\sqrt{\omega^2 + s_0^2 \cos^2(\varphi_0')}=\sqrt{\omega^2+s_0^2-s_0^2 \sin(\varphi_0)/\alpha}$ and the outgoing angle $\varphi_1$ satisfies $\sin(\varphi_1)=\omega/s_1$. Denote the speed and the angle drawn at $r_2$ by $s_2$ and $\varphi_2$.

\begin{lemma} \label{lemma: acquiring density for the disk}
Assume $|\omega-\tilde{\omega}| \geq \kappa$ for some $\kappa>0$ and that $\sin(\varphi_0) \leq \frac{\alpha \sqrt{1-\epsilon}}{\sqrt{1+\alpha^2}}$. Let $H_{\zeta_{\omega}}(\tilde{\theta}, \tilde{\omega})=(\tilde{\theta} - \zeta_{\omega},\tilde{\theta} + \zeta_{\omega}) \times (\tilde{\omega} - \zeta_{\omega},\tilde{\omega}+\zeta_{\omega})$ and let $\mu_{H_{\zeta_{\omega}}}(\tilde{\theta}, \tilde{\omega})$ be the uniform measure on $H_{\zeta_{\omega}}(\tilde{\theta}, \tilde{\omega})$ with density $1$.
Then there exists $\eta_{\omega}>0$ such that, if a particle resets $\omega$ to $\tilde{\omega}$, then $$\Pi_{\omega}[\mathcal{P}_*^{t_0+t_1} (\mu_{H_\zeta(r_0,0,s_0,\sin(\varphi_0))} \times \delta_{(\theta,\omega)})] \geq \eta_{\omega} \mu_{H_{\zeta_{\omega}}(\tilde{\theta}, \tilde{\omega})}.$$
Here $\Pi_\omega$ denotes the projection to the disk coordinates. Same for resetting from $\tilde{\omega}$ to $\omega'$
\end{lemma}

When a particle collides with the disk that rotates at a set angular velocity $\omega$, it
\textquoteleft looses' its density to the disk, i.e., $\Pi_{p_{\omega}}[\mathcal{P}_*^{t_0+t_1} (\mu_{H_\zeta(r_0,0,s_0,\sin(\varphi_0))}\times \delta_{(\theta,\omega)})]$ is not absolutely continuous. Therefore, we need to ensure that the particle \textquoteleft restores' its density once it collides with $\partial \Gamma$ again.

\begin{lemma} \label{lemma: reacquiring density}
There exists $\kappa>0$, $\eta''>0$, and $\zeta''>0$ such that if $\tilde{\omega} \in [-2\kappa,2\kappa]$ and the particle $p_{\omega}$ is sent to change the angular velocity of the disk from $\omega$ to $\tilde{\omega}$ or from $\tilde{\omega}$ to $\omega'$, then, upon return to $\partial \Gamma$ at time $t_0+t_1$, $$\Pi_{p_{\omega}}[\mathcal{P}_*^{t_0+t_1} (\mu_{H_{\zeta}(r_0,0,s_0,\sin(\varphi_0))}\times \delta_{(\theta,\omega)})] \geq \eta'' \mu_{H_{\zeta'' }(r_2,0,s_2,\sin(\varphi_2))}.$$
\end{lemma}

Combining Lemmas \ref{lemma: acquiring density}-\ref{lemma: reacquiring density} completes the proof of Prop. \ref{prop: petite}. $\square$

\medbreak

\textbf{Proof of Lemma \ref{lemma: acquiring density}: acquiring density}

Originally, each particle is assigned a point measure that evolves deterministically until collision with $\partial \Gamma$ at point $r$ and then gets perturbed in $s$ and $\sin(\varphi)$ directions. Then the measure supported on a two-dimensional sub-manifold evolves with billiard dynamics until the next collision, when it gets perturbed in $s'$ and $\sin(\varphi')$ directions. Overall, the final coordinates $(r',\xi',s'\sin(\varphi'))$ depend on the original $r$ as well as randomly drawn $s$, $\sin(\varphi)$, $s'$, and $\sin(\varphi')$. The Jacobian of this mapping does not depend on $r$ and is equal to
$$J=-\frac{\partial r'}{\partial \sin(\varphi)} \frac{\partial \xi'}{\partial s} = \frac{l(\varphi)}{\cos^2(\varphi)} \frac{l(\varphi)}{s} = \frac{l(\varphi)^2}{\cos^2(\varphi) s}$$
where $l(\varphi)$ is the distance of flight between collisions.
This determinant is clearly bounded below and above  by some $J_{\min}$ and $J_{\max}$ if $s_{\min}-\zeta \leq s \leq s_{\max}+ \zeta$ and $\alpha + \epsilon-\zeta \leq |\sin(\varphi)| \leq 1-\epsilon+\zeta$. Therefore the density in a neighborhood $H_{\zeta}(r',0, s',\sin(\varphi'))$ is bounded below by $\eta_0:=\rho_{\min}^2/ J_{\max}$ for small enough $\zeta$. $\square$

\medbreak

\textbf{Proof of Lemma \ref{lemma: pushing density forward}: Pushing density forward}

Let $t$ be the time of flight from $(r,0,s,\sin(\varphi))$ to the next collision point with $\partial \Gamma$, $(r',0,s,\sin(\varphi))$. Let us first push the measure $\mu_{H_{\zeta}(r,0,s,\sin(\varphi))}$ forward for time $t$ under the billiard flow $\mathcal{F}^t$ on the circle, i.e., the deterministic dynamics with the angles of incidence equal to the angles of reflection. Since the speed and the angle are preserved under the billiard flow, the Jacobian of $\mathcal{F}^t$ in the variables $(r,\xi,s,\sin(\varphi))$ is equal to $1$. Therefore, the density of $\mathcal{F}_*^t \mu_{H_{\zeta}(r,0,s,\sin(\varphi))}$ is equal to $1$ in the neighborhood $\mathcal{F}_*^t H_{\zeta}(r,0,s,\sin(\varphi))$, which is \textquoteleft skewed' in $r$ and $\xi$ variables. Due to bounds on $s$ and $\sin(\varphi)$, however, there exists $\zeta_1>0$ such that $H_{\zeta_1 \cdot \zeta}(r',0,s,\sin(\varphi)) \subset \mathcal{F}^t(H_\zeta(r,0,s,\sin(\varphi))$.

Pushing the measure $\mu_{H_{\zeta}(r,0,s,\sin(\varphi))}$ forward under $\mathcal{P}^t$ and keeping track only of the part that stays in a small neighborhood of $(r',0,s',\sin(\varphi'))$ is equivalent to first pushing $\mu_{H_{\zeta}(r,0,s,\sin(\varphi))}$ forward by $\mathcal{F}^t$ first (possible reflections off $L_u \cup L_d$ only change the sign of the Jacobian), then perturbing in $s$ and $\varphi$ variables, and finally applying 'the change of the chart map' from a neighborhood of $(r',0,s,\sin(\varphi))$ to a neighborhood of $(r',0,s',\sin(\varphi'))$, which essentially maps each point to where it would be if the perturbation occurred exactly at collision and not at time $t$. Indeed, the variables $s$ and $\sin(\varphi)$ are preserved under the billiard flow $\mathcal{F}$, and we can perturbing in $s$ and $\varphi$ variables is a valid operation at time $t$. The change of the chart map $T$ then fixes the gaps created by the perturbation occurring too early or too late. It maps $(\tilde{r},\tilde{\xi},\tilde{s},\sin(\tilde{\varphi}))$ from a neighborhood of $(r',0,s,\sin(\varphi))$ to $(\tilde{r'},\tilde{\xi'},\tilde{s'},\sin(\tilde{\varphi'}))$ from a neighborhood of $(r',0,s',\sin(\varphi'))$ as follows: $\tilde{r}'=\tilde{r}$, $\tilde{s}'=\tilde{s}+(s'-s)$, $\sin(\tilde{\varphi}')=\sin(\tilde{\varphi})+(\sin(\varphi')-\sin(\varphi))$, and $\tilde{\xi}' = (1 + \frac{(s'-s)}{\tilde{s}})\tilde{\xi}$, where the last relation arises from $\frac{\tilde{\xi}'}{\tilde{s}'}=\frac{\tilde{\xi}}{\tilde{s}}$.

When we perturb $\mu_{H_{\zeta_1 \cdot \zeta}(r',0,s,\sin(\varphi)) }$ in variables $s$ and $\varphi$, the fraction of the measure that stays in $H_{\zeta_1 \cdot \zeta}(r',0,s,\sin(\varphi))$ is $2\rho_{\min}(\zeta_1 \zeta)^2$, where $\rho_{\min}=\min\{\frac{4 \beta^{3/2}}{\sqrt{\pi}}s_{\min}^2 e^{-\beta s_{\min}^2},\frac{4 \beta^{3/2}}{\sqrt{\pi}}s_{\max}^2 e^{-\beta s_{\max}^2}\}$. The Jacobian for the change of the chart mapping $T$ is $(1 + \frac{(s'-s)}{s})$, which greater or equal to $\frac{s_{\min}-\zeta}{s_{\max}-\zeta}$. In addition, $H_{\frac{s_{\min}-\zeta}{s_{\max}-\zeta}\zeta_1\zeta}(r',0,s',\sin(\varphi')) \subset T H_{\zeta_1 \cdot \zeta}(r',0,s,\sin(\varphi))$. Therefore,
$$\mathcal{P}^t \mu_{H_\zeta(r,0,s,\sin(\varphi'))} \geq 2 \rho_{\min} (\zeta_1 \zeta)^2  \frac{s_{\min}}{s_{\max}} \mu_{H_{\frac{s_{\min}-\zeta}{s_{\max}-\zeta}\zeta_1\zeta}(r',0,s',\sin(\varphi'))}$$

Lemma \ref{lemma: pushing density forward} follows. $\square$

\medbreak

\textbf{Proof of Lemma \ref{lemma: acquiring density for the disk}: acquiring density for the disk.}
After collision, the disk variables satisfy $\tilde{\omega}=s_0 \sin(\varphi_0)/\alpha$ and $\tilde{\theta}=\theta_0 + \omega t_{\mathrm{col}}+ \tilde{\omega}(t-t_{\mathrm{col}})=\theta_0 + \tilde{\omega} t + (\omega-\tilde{\omega})t_{\mathrm{col}}$, where $t_{\mathrm{col}}=\frac{l(\varphi_0)}{s_0}$ is the collision time of the particle and $t=t_0+t_1$. An easy computation shows that the Jacobian for this mapping is
\begin{equa}
J_{\omega} & =\frac{(\omega-\tilde{\omega})}{\alpha s_0} l(\varphi_0)[\cos(\varphi_0)-\sin(\varphi_0) \frac{\partial l(\varphi_0)}{\partial \varphi_0}] \\
& = \frac{(\omega-\tilde{\omega})}{\alpha s} l(\varphi_0)[\cos(\varphi_0)-\sin(\varphi_0)\tan(\varphi_0')].
\end{equa}
If we choose $|\sin(\varphi_0)|<\frac{\alpha \sqrt{1-\epsilon}}{\sqrt{1+\alpha^2}}$ and $|\omega - \tilde{\omega}| \geq \kappa$, then there are uniform lower ($J^\omega_{\min}$) and upper ($J^\omega_{\max}$) bounds on the Jacobian and there exists $\zeta_{\omega}$ such that the pushed forward measure has density at least $1/J^{\omega}_{\max}$ in a neighborhood $(\tilde{\theta}-\zeta_{\omega},\tilde{\theta}+\zeta_{\omega})\times(\tilde{\omega}-\zeta_{\omega},\tilde{\omega}+\zeta_{\omega})$. $\square$

\medbreak

\textbf{Proof of Lemma \ref{lemma: reacquiring density}: re-acquiring density for the particle that hit the disk. }
Similarly the Jacobian for the mapping from $(s,\sin(\varphi),s',\sin(\varphi'))$ to $(r',\xi',s',\sin(\varphi'))$ can be computed given that the particle resets the disk from $\omega$ to $\omega'$. If we plug either $\omega=0$ or $\omega'=0$, the Jacobian is bounded above and below by some positive constants. Therefore, there exists $\kappa>0$ such the Jacobian for resetting to/from any $\tilde{\omega} \in [-2\kappa,2\kappa]$ will also have lower and upper bounds. By the same reasoning as in Lemma \ref{lemma: acquiring density}, we conclude that there exist $\eta''>0$, and $\zeta''>0$ such that $\Pi_{i}[\mathcal{P}_*^{t_0+t_1} (\mu_{H_{\zeta}(r_0,0,s_0,\sin(\varphi_0))}\times \delta_{(\theta,\omega)})] \geq \eta'' \mu_{H_{\zeta''}(r_2,0,s_2,\sin(\varphi_2))}$. $\square$

\subsection{Mixing} \label{subsect: mixing}

Mixing and convergence of initial distributions to the invariant measure follow almost immediately from our existence argument for the invariant probability measure. In the proof of Prop.~\ref{prop: petite}, a lot of effort has been devoted to guarantee lower bounds on the densities of the pushed forward measures. A weaker property of Markov processes, irreducibility, can be shown in a similar manner by dropping the lower bounds on times and densities and allowing the paths we follow to start anywhere in the phase space.

\begin{definition}
A continuous-time Markov process $\Phi_t$ is called irreducible if for all $\mathbf{z} \in \Omega$, whenever Leb$(A)>0$, there exists some $t>0$, possibly dependent on both $\mathbf{z}$ and $A$, such that $\mathcal{P}^t(\mathbf{z},A)>0$.
\end{definition}

\begin{lemma} \label{lemma: irreducible}
Markov process $\Phi_t$ is irreducible.
\end{lemma}

The proof of Lemma \ref{lemma: irreducible} is a simple modification of the proof of Prop. \ref{prop: petite}.

\begin{definition}
The Markov process $\Phi_t$ is called ergodic if an invariant probability measure $\mu$ exists and
$$\lim\limits_{t \to \infty} \|\mathcal{P}^t(\mathbf{z},\cdot)-\mu\|=0, \;\;\;\;\; \forall \mathbf{z} \in \Omega,$$
where $\|\cdot\|$ is the total variation norm.
\end{definition}

The proof of ergodicity of $\Phi_t$ relies on sampling the Markov process $\Phi_t$ at integer times, which generates a discrete-time skeleton chain $\Phi_1$.  Denote the transition probability kernel for $\Phi_1$ by $\mathcal{P}_1$. The following theorem by Meyn and Tweedie relates skeleton chains to the ergodicity of the Markov processes.

\begin{theorem} \cite[Thm.~6.1]{MeynII} \label{thm: Meyn ergodicity}
Suppose the Markov process $\Phi_t$ is irreducible and $\mu$ is an invariant probability measure for $\Phi_t$. Then $\Phi_t$ is ergodic if and only if $\Phi_1$ is irreducible.
\end{theorem}

\begin{proposition} \label{prop: ergodic}
The Markov process $\Phi_t$ is ergodic.
\end{proposition}

The mixing of the invariant measure for the Markov process $\Phi_t$ and the convergence of initial distributions to the invariant measure follow from ergodicity by the Dominated Convergence Theorem.
Indeed,
\begin{equa}
\lim\limits_{t \to \infty}\|\mathcal{P}^t\nu - \mu\| & =\lim\limits_{t \to \infty} \sup\limits_{A \subset \Omega}|\int_\Omega (\mathcal{P}^t(
\mathbf{z},A)-\mu(A))d\nu| \\
& \leq \lim\limits_{t \to \infty} \int_\Omega \|P^t(\mathbf{z},\cdot)-\mu(\cdot)\|d\nu= 0,
\end{equa}
and to show mixing one may replace $\Omega$ with $B$ in the second integral.

\medbreak

\textbf{Proof of Prop.~\ref{prop: ergodic}.}

Ergodicity follows once we show that the time-$1$ sampled chain $\Phi_1$ is irreducible. The proof is a modification of the proof of Prop.~\ref{prop: petite}. What we really need to show is that for any $\mathbf{z},\mathbf{z}' \in \Omega$, there exists $T \in \mathbb{N}$ and a sample path $\sigma(\mathbf{z},\mathbf{z}',T)$ from $\mathbf{z}$ to $\mathbf{z}'$ in time $T$. In addition, some density is acquired and carried through along the path. No lower bounds are required. The existence of a sample path is guaranteed in the proof of Lemma \ref{lemma: path} noting that allowing $\mathbf{z}$ and $\mathbf{z}'$ to be not in $C$ only changes times it takes for particles to reach $\partial \Gamma$ for the first time in forward or backward times respectively. Time $\tilde{T}$ for Lemma \ref{lemma: path subproblem} is allowed to vary in a range of values. In particular, we can choose $\tilde{T}$ such that $T$ is integer (here it does not have to be the same for all states either). Modifying Lemmas \ref{lemma: acquiring density}-\ref{lemma: reacquiring density} to apply for all injection parameters and dropping the lower bounds, we conclude that $\Phi_1$ is irreducible. $\square$

\section{Sub-exponential mixing} \label{sect: non exp mixing}

\begin{proposition} \label{prop: non exp mixing}
There exist (many) initial probability distributions $\lambda$ on $\Omega$ that converge to the unique invariant measure $\mu$ with sub-exponential rates, i.e. $\exists \varsigma>0$ such that for $T$ large enough
$$\|\mathcal{P}^T_* \lambda - \mu\| \geq \frac{\varsigma}{T^2}.$$
In particular, the unique invariant measure $\mu$ for the Markov Process $\Phi_t$ is not exponentially mixing, i.e. there exist a Borel set $A \subset \Omega$ such that
$$\sup\limits_{B \in \mathcal{B}}|\int\limits_{A}\mathcal{P}^T(\mathbf{z},B)d \mu - \mu(A)\mu(B)| \geq \mu(A) \times \frac{\varsigma}{T^2}.$$
\end{proposition}

\textbf{Proof of Prop.~\ref{prop: non exp mixing}.}

Let $B_T$ be a set of $\mathbf{z} \in \Omega$ such that at least one particle in $\mathbf{z}$ does not have any collisions for time $T$. We are interested in estimating $\mu(B_T)$.

The proof is rather similar to the sub-exponential mixing proof for a system driven by thermostats in \cite{Y2} except the dynamics on $B_T$ is not deterministic and there is no potential to aid the estimates on an upper bound on the measure of $B_T$.

Prop.~\ref{prop: petite} ensures that there exist $T_0, \eta>0$ such that $\mathcal{P}^{T_0}(x,\cdot) \geq \eta m_C$, where $\eta>0$ and $m_C$ is the uniform probability measure on $C$. By the invariance of $\mu$, we conclude $\mu \geq \mu(C) \eta m_C$.

Let $\delta<d/(s_{\max}\sqrt{\frac{\alpha^2-\epsilon+2}{\alpha^2+1}})$. This guarantees that for any $\mathbf{z} \in C$, any particle in $\mathbf{z}$ that experiences a collision with $\partial \Gamma$ at time $t \leq \delta$ does not experience any disk collisions on time interval $[0,t)$.

Denote by $\Omega_i$ the projection of the phase space $\Omega$ into components associated with $i^{th}$ particle and let $m_{\Omega_i}$ be the uniform probability measure on $\Omega_i$. Let $G_i=\{(r_i,s_i,\varphi_i,\xi_i):$ the particle will collide with $\partial \Gamma$ in time $\delta \}$. Here we use coordinates defined in subsect.~\ref{subsect: tau}.  Let $\gamma=m_{\Omega_i}(G_i)>0$. Then if $\mathbf{z}$ is drawn uniformly in $C$, with probability $1-(1-\gamma)^k$, at least one particle in $\mathbf{z}$ will hit $\partial \Gamma$ in time $\delta$.

Suppose we start with a particle that collides with $\partial \Gamma$ at time $t \leq \delta$. Then we are interested in the lower bound on the probability that,  once we draw its new $s$ and $\varphi$, it will not collide with the disk or $\partial \Gamma$ in time $T+(\delta - t)$. By replacing $T + (\delta - t)$ with $T + \delta$ we only lower such probability. In addition, consider only the situations when a particle will fly at least distance $d$ before the next collision, i.e. $|\sin(\varphi)| \leq \sqrt{1-(\frac{d/2}{R+d})^2}$, which happens with probability $\sqrt{1-(\frac{d/2}{R+d})^2}$. Then the probability that a randomly emitted particle will not collide with the disk or $\partial \Gamma$ in time $T+\delta$ is greater or equal than

$$\sqrt{1-(\frac{d/2}{R+d})^2} \int_0^{\frac{d}{T+\delta}} \frac{4 \beta^{3/2}}{\sqrt{\pi}}s^2 e^{-\beta s^2} ds $$
$$= \sqrt{1-(\frac{d/2}{R+d})^2} -2\sqrt{\frac{\beta}{\pi}}\frac{d}{T+\delta}e^{-\frac{\beta d^2}{(T+\delta)^2}}+\mathrm{Erf}[\frac{\sqrt{\beta}d}{T+\delta}]$$
$$=\sqrt{1-(\frac{d/2}{R+d})^2} \frac{4}{3}\frac{\beta^{3/2}}{\sqrt{\pi}} \frac{d^2}{(T+\delta)^3}+O(\frac{1}{T^5}) \geq \frac{\sigma}{T^3}$$

for some $\sigma>0$ and $T$ large enough.

Then at least $(1-(1-\gamma)^k) \times \frac{\sigma}{T^3}$-fraction of $m_C$, and therefore at least $\mu(C) \eta (1-(1-\gamma)^k) \times \frac{\sigma}{T^3}$ fraction of $\mu$, ends up in $B_T$ in time $\delta$. By the invariance of $\mu$, we conclude that $\mu(B_T)>\mu(C) \eta (1-(1-\gamma)^k) \times \frac{\sigma}{T^3}$ for large enough $T$.

\medbreak

To obtain an upper bound on the fraction of $\mu$ that will end up in $B_T$, we observe that the only ways to get to $B_T$ not from $B_T$ are to emit a slow enough particle from $\partial \Gamma$ or to acquire slow enough speed after a disk collision. In the later case, the speed after collision $\sqrt{\omega^2 + s^2\cos^2(\varphi')}$ must be less or equal to $\frac{l}{T}$, where $l$ is the maximal distance of flight from $\partial D$ to $\partial \Gamma$. In particular, $s\cos(\varphi')= s\sqrt{1- \sin^2(\varphi)/\alpha^2}$ must be less or equal to $\frac{l}{\Gamma}$, i.e. $s \leq \frac{\mathrm{c}}{T}$ for some $c>0$. Therefore, starting from any initial distribution, the probability to end up in $B_T$ is bounded above by the probability of drawing $s < \frac{\max\{c,1\}}{T}$, which is equal to

$$\int_0^{\frac{\max\{c,1\}}{T}} \frac{4 \beta^{3/2}}{\sqrt{\pi}}s^2 e^{-\beta s^2} ds \leq \frac{\sigma'}{T^3}$$

for some $\sigma'>0$ and $T$ large enough.

Thus we obtain that for any $n \in \mathbb{N}$ and $T$ large enough, $\mu(B_{T})=\mu(B_{T+n\delta})+nf(T)$, where $f(T) \approx \frac{1}{T^3}$. Here $f \approx g$ means $f(T)=\Theta(g(T)$, i.e. there exists $\xi, \xi'>0$ such that $\xi' g(T) \leq f(T) \leq \xi g(T)$.

Summing up, we conclude that $\mu(B_T) \approx \sum\limits_{k=0}^\infty \frac{\xi}{(T+k\delta)^3} \approx \frac{1}{T^2}$.

Note that the dynamics for our system is statistically very similar to the dynamics of an expanding map with a neutral fixed point (aka the Pomeau-Manneville map). Indeed, the mass originally in $B_T$ evolves in $B_T$ for at least time $T$; extra mass is \textquoteleft deposited' from the parts of the phase space where particles experience collisions. One can complete the proof of Prop. \ref{prop: non exp mixing} using an argument very similar to \cite{Y2} and \cite{LSY}.

Let $\lambda \ll \mu$ with $d\lambda = \varphi d\mu$ be such that $\varphi \geq 1+c$ on $B_{T_0}$ for some small $c>0$ and any $T_0$.
Then for $k\delta>T_0$

\begin{equa}
\|\mathcal{P}^n\delta_* \lambda - \mu\| & \geq (\mathcal{P}^n\delta_* \lambda)(B_{k\delta}) - \mu(B_{k\delta}) \geq \lambda(B_{k\delta+n\delta})-\mu(B_{k\delta})\\
       & \geq (1+c)\mu(B_{k\delta+n\delta})-\mu(B_{k\delta}) = [(1+c)\frac{\mu(B_{k\delta+n\delta})}{\mu(B_{k\delta})}-1]\mu(B_{k\delta}) \\
       & =[(1+c)\frac{\mu(B_{k\delta+n\delta})}{\mu(B_{k\delta+n\delta})+n f_n(k\delta)}-1]\mu(B_{k\delta}) \\
       & =[c-(1+c)\frac{n f(k \delta)}{\mu(B_{k\delta+n\delta})+n f(k\delta)}]\mu(B_{k\delta})
\end{equa}

Then for fixed $N$ large enough and $k=Nn$
       $$\|\mathcal{P}^n\delta_* \lambda - \mu\| \geq [c-(1+c)\frac{n f(Nn
       \delta)}{\mu(B_{(N+1)n\delta})+n f(Nn\delta))}]\mu(B_{Nn\delta})
       \geq \frac{c}{2}\frac{\xi'}{N^2 n^2 \delta^2}.$$

\medbreak
To obtain a lower bound on the rate of mixing, it suffices to pick $A=B_{T_0}$ and $\lambda \ll \mu$ such that $d\lambda = (\mathbf{1}_A/\mu(A))d\mu$. Then
$$\sup\limits_{B \in \mathcal{B}}\frac{1}{\mu(A)}|\int\limits_{A}\mathcal{P}^T(\mathbf{z},B)d\mu-\mu(A)\mu(B)| = \|\mathcal{P}^{T}_* \lambda - \nu\| \geq \frac{\varsigma}{T^2}.$$
\begin{flushright}
$\square$
\end{flushright}

\textbf{Acknowledgements:}\\
The author would like to thank the unnamed referees for valuable comments and suggestions, Lai-Sang Young for enlightening discussions; Jean-Pierre Eckmann and Noe Cuneo for valuable comments. The author was supported by the ERC Advanced Grant "Bridges"
\medbreak

\end{document}